\documentclass[journal,comsoc,final,letterpaper,twoside,twocolumn,romanappendices]{IEEEtran}
\usepackage{silence} 
\usepackage[T1]{fontenc}
\usepackage[latin9]{inputenc}
\usepackage{amsmath,amsfonts,amsthm,amssymb}
\usepackage{algorithmic,algorithm}
\usepackage{array,stackrel}
\usepackage[caption=false,font=normalsize,labelfont=sf,textfont=sf]{subfig}
\usepackage{textcomp}
\usepackage{stfloats}
\usepackage{url}
\usepackage{verbatim}
\usepackage{graphicx}
\usepackage{cite}
\usepackage{xcolor}
\usepackage{pdfcolmk}
\usepackage{refcount}
\usepackage{esint}
\usepackage{hhline}
\usepackage{bm}
\usepackage{xfrac}
\usepackage{makecell}
\usepackage{float}
\usepackage{cases}
\usepackage[]{footmisc}

\usepackage{booktabs}
\PassOptionsToPackage{normalem}{ulem}
\usepackage{ulem}
\usepackage[unicode=true,
 bookmarks=true,bookmarksnumbered=true,bookmarksopen=true,bookmarksopenlevel=1,
 breaklinks=false,pdfborder={0 0 0},pdfborderstyle={},backref=false,colorlinks=false]
 {hyperref}
\hypersetup{pdftitle={Your Title},
 pdfauthor={Your Name},
 pdfpagelayout=OneColumn, pdfnewwindow=true, pdfstartview=XYZ, plainpages=false}
\PassOptionsToPackage{unicode}{hyperref}
\PassOptionsToPackage{naturalnames}{hyperref}
\usepackage[cmintegrals]{newtxmath}
\usepackage[caption=false,font=normalsize,labelfont=sf,textfont=sf]{subfig}
\newcounter{MYtempeqncnt}

\makeatletter

\let\oldforeign@language\foreign@language
\DeclareRobustCommand{\foreign@language}[1]{%
  \lowercase{\oldforeign@language{#1}}}
\theoremstyle{plain}
\newtheorem{thm}{\protect\theoremname}
\theoremstyle{plain}

\usepackage{stackengine}

\usepackage[caption=false,font=normalsize,labelfont=sf,textfont=sf]{subfig}

\makeatother

\providecommand{\lemmaname}{Lemma}
\providecommand{\theoremname}{Theorem}

\usepackage{tabularx,booktabs}
\newcolumntype{C}{>{\centering\arraybackslash}X} 
\begin{document}
\title{Closed-Form and Asymptotic BER Analysis of the Fluctuating Double-Rayleigh with Line-of-Sight Fading Channel}

\author{Aleksey S.~Gvozdarev\href{https://orcid.org/0000-0001-9308-4386}{\includegraphics[scale=0.1]{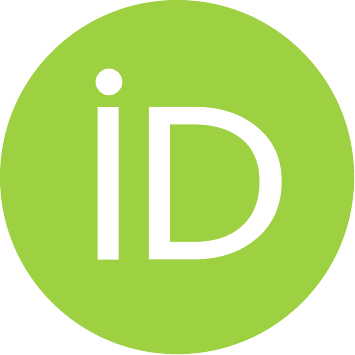}},~\IEEEmembership{Member,~IEEE,}
\thanks{The author is with the Department of Intelligent Radiophysical  Information Systems, P. G. Demidov Yaroslavl State University, 150003 Yaroslavl,
Russia (e-mail: asg.rus@gmail.com).}
\thanks{\copyright 2022 IEEE.  Personal use of this material is permitted.  Permission from IEEE must be obtained for all other uses, in any current or future media, including reprinting/republishing this material for advertising or promotional purposes, creating new collective works, for resale or redistribution to servers or lists, or reuse of any copyrighted component of this work in other works.}}



\maketitle

\begin{abstract}
Recently, a generalization of the double-Rayleigh with line-of-sight channel fading model taking into account shadowing of the line-of-sight component has been proposed. In this research, a closed-form analysis of the average bit error rate for MPSK/MQAM modulations is performed. The derived solution is accompanied by proposed numerically efficient approximation and all possible asymptotic expressions that correspond to extreme channel parameters. Lastly, a numerical simulation was performed that demonstrated the correctness of the derived results.
\end{abstract}


\section{Introduction}
\IEEEPARstart{N}{owadays}, the communication link quality of the modern ad-hoc  systems is mainly limited by the wireless signal propagation effects. Thus the choice of channel model heavily impacts the predicted overall system performance.

Recently, a novel fluctuating double-Rayleigh with line-of-sight (fdRLoS) fading channel model was proposed \cite{Lop21}. It generalizes the double-Rayleigh with line-of-sight (LoS) model \cite{Sal06} by including the Gamma-distributed shadowing of the LoS component, and covers such physical scenarios as the  "pipe-like"/keyhole channel \cite{Ges02},  propagation via diffracting street corner \cite{Erc97}, amplify-and-forward relay \cite{Has04}, free-space optical communication through a turbulent medium \cite{And85} and Vehicle-to-vehicle (V2V) communications \cite{Ai18}.

The original work by Lopez-Fernandez et al. \cite{Lop21} states the results for the probability density function (PDF) and cumulative distribution function (CDF), expressed for integer values of LoS shadowing parameter; and outage probability, derived via the obtained CDF.  The problem with the obtained expressions is that they are formularized in a way that does not make further analytical derivations possible. Moreover, for similar models, experimental results demonstrate that the shadowing parameter efficiently can be non-integer and generally less than 1 (including the so-called hyper-Rayleigh regime \cite{Gar19}), which can not be deduced from the results in \cite{Lop21}. Furthermore, the average bit error rate (ABER) analysis quantifying the communication link quality for such a channel is not present.

Motivated by the problems stated above, in this letter a closed-form expression for the ABER is derived for the fdRLoS channel model with arbitrary parameters. It is succeeded with the computationally efficient approximation obtained by truncating the derived solution, and the truncation error is estimated. Then the asymptotic expressions of the ABER for all possible extreme cases are evaluated: (\textit{a}) high signal-to-noise ratio (SNR), (\textit{b}) extremely high/low shadowing; and (\textit{c}) extremely strong/weak LoS component. Finally, to validate the correctness and accuracy of the analytical work, computer simulation was performed, and the obtained numerical results were studied. It was found out that specifically for the hyper-Rayleigh regime, the minimum ABER is achieved when the total power of multipath components equals the power of LoS component.

\section{Preliminaries}
\subsection{Channel model: physical and statistical description}
Let us start with a brief description of the proposed in \cite{Lop21} fluctuating double-Rayleigh with line-of-sight fading channel model. It is generally assumed that the signal propagating within the wireless channel  can be represented as the combination of the line-of-sight component that undergoes shadowing with average magnitude $\omega_0$ and uniformly distributed phase $\phi                \sim U[0,2\pi )$ (for conformity one uses the initial notation given in \cite{Lop21}) and double-Rayleigh fading (dRf) component $\omega_2 G_2 G_3$:
\begin{IEEEeqnarray}{rCl}\label{channel-model}
S=\omega_0\sqrt{\xi }e^{j\phi }+\omega_2 G_2 G_3.
\end{IEEEeqnarray}
Here $\xi$ is the shadowing parameter following Gamma distribution normalized to have unit power and shape coefficient $m$; $\omega_2$ is the average magnitude of the fluctuating double-Rayleigh component; and $G_2, G_3$ are the zero-mean complex normal random variables (i.e., $G_2, G_3                                 \sim \mathcal{CN}(0,1)$).

\begin{figure*}[!b]
\hrulefill
\normalsize
\setcounter{MYtempeqncnt}{0}
\setcounter{equation}{4}
\vspace*{-4pt}
\begin{IEEEeqnarray}{rCl}\label{thm-1}
&&J_Q=\frac{\sin(\hat{m}\pi )}{4\pi }\sum_{l=0}^{\infty}\sum_{n=0}^{\infty}\frac{\left(\frac{1}{2}\right)_{l+n}}{(2)_{l+n}}\frac{\left(\frac{\hat{m}(K+1)}{\bar{\gamma }\delta_{2,j}K+\hat{m}(K+1)}\right)^{n+\hat{m}}}{l!n!}  G_{0,1:1,1:1,1}^{1,0:1,1:1,1}\left(
\begin{array}{c}
\mbox{---}\\
0\\
\end{array}\middle\vert
\begin{array}{c}
1-\hat{m}-n\\
0\\
\end{array}\middle\vert
\begin{array}{c}
\hat{m}-l \\
0\\
\end{array}\middle\vert
\frac{\hat{m}\bar{\gamma }\delta_{2,j}}{\bar{\gamma }\delta_{2,j}K+\hat{m}(K+1)},\frac{\bar{\gamma }\delta_{2,j}}{K+1}
\right).
\end{IEEEeqnarray}

\vspace*{-4pt}
\begin{IEEEeqnarray}{rCl}\label{cor-1}
\hspace{-5pt}{\rm ABER_{err}}(L,N)\leq \delta_{1}\sum_{j=1}^{\delta_{3}}\frac{\frac{K+1}{\bar{\gamma }\delta_{2,j}}e^{\frac{K+1}{\bar{\gamma }\delta_{2,j}}}}{4\left(\frac{K+1}{\bar{\gamma }\delta_{2,j}}+\frac{K}{m}\right)^{m+N}}\Gamma \left(m-L,\frac{K+1}{\bar{\gamma }\delta_{2,j}}\right)\left|\mbox{}_2F_1\left(\frac{1}{2},1;2;\frac{K+1}{\bar{\gamma }\delta_{2,j}}\right)-
\sum_{l=0}^{L}\sum_{n=0}^{N}\frac{(\sfrac{1}{2})_{l+n}(1-m)_l(m)_n}{(2)_{l+n}}\frac{\left(\frac{K+1}{\bar{\gamma }\delta_{2,j}}\right)^{l+n}}{n!l!} \right|
\end{IEEEeqnarray}

\end{figure*}

Assuming that the channel is normalized (i.e., $\mathbb{E}{|S|^2}=1$) and noticing that $|G_2|^2$ follows exponential distribution with unit mean value (see \cite{Lop21}), the probability density function of the instantaneous signal-to-noise ratio $\gamma$ (defined in terms of the average signal-to-noise ratio $\bar{\gamma }$ and the squared signal envelope $|S|^2$, i.e., $\gamma=\bar{\gamma }|S|^2$) is given by (see \cite{Lop21})
\setcounter{equation}{1}
\begin{IEEEeqnarray}{rCl} \label{eq-pdf-full}
&&f_{\gamma }(\gamma )=\int_{0}^{\infty}f_{\gamma_x}(\gamma|x) e^{-x}{\rm d}x
\end{IEEEeqnarray}
where $f_{\gamma_x}(\gamma|x)$ is the conditional probability density function conditioned on $x=|G_3|^2$ and defined as
\begin{IEEEeqnarray}{rCl} \label{eq-pdf-conditional}
&&\hspace{-10pt}f_{\gamma_x}(\gamma|x)=\frac{m^m(1+k_x)}{(m+k_x)^m\bar{\gamma }_x}e^{-\frac{1+k_x}{\bar{\gamma }_x}\gamma }\!\! \mbox{}_1F_1\left(m,1,\!\frac{k_x(1+k_x)}{k_x+m}\frac{\gamma }{\bar{\gamma }_x}\right),
\end{IEEEeqnarray}
where $\mbox{}_1F_1(                                \cdot )$ denotes the confluent hypergeometric function \cite{DLMF}, $\bar{\gamma }_x=\frac{K+x}{K+1}\bar{\gamma }$, $k_x=\frac{K}{x}$, $K=\frac{\omega_0^2}{\omega^2_2}$ is the Rician K-factor, and  $m$ is responsible for LoS shadowing intensity.

It should be specifically emphasized that due to the complex substitutions, \eqref{eq-pdf-full} does not have a closed-form solution for arbitrary values of parameters, although \cite{Lop21} presents a simplified result for the case of $m                \in \mathbb{Z}_{+}$. Its weak point is that it does not cover a practically highly valuable case of $0.5\leq m<1$, which constitutes the heaviest fading scenario (the so-called hyper-Rayleigh, see \cite{Fro07, Gar19}).
\subsection{System performance metrics}
The primary metric, assumed herein to characterize wireless communication link quality in the presence of fading, is the average bit error rate. It is defined in terms of the instantaneous BER (i.e., $\mathrm{BER}\left(\gamma \right)$) averaged over the stochastic variations of the instantaneous signal-to-noise ratio with the PDF $f_{\gamma }\left(\gamma \right)$:
\begin{IEEEeqnarray}{rCl} \label{eq:ABER}
&&\mathrm{ABER}=\delta_{1}\sum_{j=1}^{\delta_{3}}\int_{0}^{\infty}Q(\sqrt{2\delta_{2,j}\gamma })f_{\gamma }(\gamma ){\rm d}\gamma,
\end{IEEEeqnarray}
with $Q(              \cdot )$ being the Gauss Q-function. If should be noted that \eqref{eq:ABER} is the so-called ``BER unified approximation`` (see \cite{Lu99}) and holds true for a wide variety of modulation schemes with the set of coefficients $\left\{ \delta_{1},\delta_{2,j},\delta_{3}\right\} $ explicitly defined for specific modulation 
(see, for instance, \cite{Sim05, Lu99}): for M-QAM $\left\{ \frac{4\left(1-\sfrac{1}{\sqrt{M}}\right)}{\log_2 M}, \frac{3(2j-1)^2}{2(M-1)}, \frac{\sqrt{M}}{2}\right\}$, for M-PSK $\left\{ \frac{1}{\max(2,\log_2 M)}, 2\sin^2\left(\frac{(2j-1)\pi }{M}\right), \max\left(1,\frac{M}{4}\right) \right\}$.

Thus the problem of the closed-form analytical ABER description efficiently boils down to the solution of the integral $J_Q=\int_{0}^{\infty}Q(\sqrt{2\delta_{2,j}\gamma })f_{\gamma }{\rm d}\gamma$.

Even though the problem is typical for such a formulation and has been numerously studied \cite{Sim98,Sim05,Gvo21,Sou13,Odr09}, the availability of the close-form solution highly depends on the form of $f_{\gamma }(\gamma )$.  It must be pointed out that due to the novelty of the assumed model, the solution of $J_Q$ up to now doesn't exist, and because of the discussed earlier complexity of the PDF \eqref{eq-pdf-full} cannot be directly obtained from the existing results. Moreover, if numerical integration in $J_Q$ is utilized, this procedure is highly sensitive to the parameter values, and for large $m$ or $k$ is unstable, which means that higher working accuracy and precision make the solution time-consuming.

\section{Derived results}
Let us derive the closed-form solution for the integral $J_Q$ by using the moment generating function (MGF) approach. First, a conditional MGF for the PDF \eqref{eq-pdf-conditional} is evaluated (valid for arbitrary $m$), and the conditional version of $J_Q$ is obtained, which is further averaged with the help of \eqref{eq-pdf-full}.
The result is given by the following Theorem 1.
\begin{thm}
For the fluctuating double-Rayleigh with line-of-sight fading channel the following statements hold true:
\begin{itemize}[
    \setlength{\IEEElabelindent}{\dimexpr-\labelwidth-\labelsep}
    \setlength{\itemindent}{\dimexpr\labelwidth+\labelsep}
    \setlength{\listparindent}{\parindent}
  ]
  \item the closed-form expression of the integral  $J_Q$ for arbitrary positive values of the shadowing parameter $m$ is given by \eqref{thm-1} (see at the bottom of the page), where $G_{0,1:1,1:1,1}^{1,0:1,1:1,1}(              \cdot )$ is the extended generalized bivariate Meijer G-function EGBMG\footnote{
EGBMG is defined in terms of the double Mellin-Barnes integral (see equation (13.1) in \cite{Hai92}) with the integration contours $\mathcal{L}_s, \mathcal{L}_t$ in the corresponding domains of complex variable $s$ and $t$ chosen in such a way to separate the specific singularities of the integrand. For practical implementation, computation methods and procedures see \cite{Ans11,Gar14}} and $\hat{m}=\begin{cases}m,m                \notin                 \mathbb{Z}^{+}_0 \\ m(1+\Delta ), m                \in \mathbb{Z}^{+}_0\end{cases}$, with $\Delta$ being the infinitesimal shift;
  \item the computationally efficient approximation of the ABER \eqref{eq:ABER} is given by $\mathrm{ABER}\approx\delta_{1}\sum_{j=1}^{\delta_{3}}J_Q(L,N)$, where $J_Q(L,N)$ is the truncated version of \eqref{thm-1} with $(L,N)$ remaining terms, and the induced truncation error (${\rm err(L,N)}=J_Q-J_Q(L,N)$) is upper-bounded by \eqref{cor-1} (see at the bottom of the page).
\end{itemize}

\end{thm}
\begin{IEEEproof}For proof see APPENDIX I.\end{IEEEproof}

The formulated results help to form a solid ground for further closed-form analysis as well as numerical optimization of the wireless communication system performance functioning  in the presence of fdRLoS channels. From the practical point of view, the proposed approximation is given in terms of the double series the converge very quickly, and for moderate $m$, even a single term is enough to deliver at least $3$-digit accuracy (see Section IV for numerical examples).

Moreover, in real-life applications, it is important to understand to what extent the channel impacts ABER. This can be estimated by evaluating the performance for the extreme fading conditions, i.e., for all possible range of channel parameters, which is given by the following Theorem 2.
\begin{thm}
In the extreme cases, the integral $J_{Q}$, defining the limiting performance of the assumed modulation schemes  for the fdRLoS channel, is given by:
\begin{itemize}[
    \setlength{\IEEElabelindent}{\dimexpr-\labelwidth-\labelsep}
    \setlength{\itemindent}{\dimexpr\labelwidth+\labelsep}
    \setlength{\listparindent}{\parindent}
  ]
  \item in the case of the high SNR regime (i.e., $\bar{\gamma }\to\infty$)
  \begin{IEEEeqnarray}{rCl} \setcounter{equation}{7}\label{thm-2-1}
&&J_Q\big|_{\bar{\gamma }\to \infty}                \sim \frac{(K+1)}{4\bar{\gamma }\delta_{2,j}}\Gamma (m)U\left(m,1\frac{K}{m}\right),
\end{IEEEeqnarray}
where $U(                         \cdot )$ is the Tricomi confluent hypergeometric function;
  \item in the case of the strong dominant component (i.e., $K\to\infty$)
  \begin{IEEEeqnarray}{rCl}\label{thm-2-2}
&&\hspace{-10pt}J_Q\big|_{K \to\infty}                \sim \frac{1}{2\sqrt{\pi }}\frac{\Gamma \left(m+\frac{1}{2}\right)}{\Gamma (m+1)}\frac{\mbox{}_2F_1\left(\frac{1}{2},m;m+1;\frac{m}{m+\bar{\gamma }\delta_{2,j}}\right)}{\left(1+\frac{\bar{\gamma }\delta_{2,j}}{m}\right)^m},
\end{IEEEeqnarray}
where $\mbox{}_2F_1(                         \cdot )_l$ is the Gauss hypergeometric function;
  \item in the case of the weak dominant component (i.e. $K\to 0$)
  \begin{IEEEeqnarray}{rCl}\label{thm-2-3}
\hspace{-30pt}J_Q\big|_{K\to 0}&\sim& \frac{1}{2}-\frac{\sqrt{\pi }}{4}U\left(\frac{1}{2},0\frac{1}{\bar{\gamma }\delta_{2,j}}\right)=\\
\hspace{20pt}&=&\frac{1}{2}-\frac{e^{\frac{1}{2\bar{\gamma }\delta_{2,j}}}}{4\bar{\gamma }\delta_{2,j}}\left(K_1\left({\frac{1}{2\bar{\gamma }\delta_{2,j}}}\right)-K_0\left({\frac{1}{2\bar{\gamma }\delta_{2,j}}}\right)\right)\IEEEnonumber,
\end{IEEEeqnarray}
where $K_0(                \cdot ), K_1(                \cdot )$ are the modified Bessel functions;
 \item in the case of the weak shadowing (i.e. $m\to \infty$) and heaviest shadowing  (i.e. $m\to \sfrac{1}{2}$)
  \begin{IEEEeqnarray}{rCl}\label{thm-2-4}
&&\hspace{-10pt}  J_Q\big|_{m \to \sfrac{1}{2}}=J_{Q}(L,N,m=\sfrac{1}{2}), \quad J_Q\big|_{m \to\infty}                \sim J_{Qm \infty},
\end{IEEEeqnarray}
where $J_{Q}(L,N,m=\sfrac{1}{2})$ is defined by the truncated version of \eqref{thm-1} and $J_{m\infty}$ is defined in \eqref{eq_thm-2-5}.
\end{itemize}
\end{thm}
\begin{IEEEproof}For proof see APPENDIX II.\end{IEEEproof}

It should be noted that all of the special functions used in \eqref{thm-2-1}-\eqref{thm-2-4} are readily accessible in all modern computer algebra systems for further numeric and analytical computations.

To the best of the author's knowledge, the ABER analysis of the fdRLoS channel is absent in current scientific literature and the results \eqref{thm-1}-\eqref{thm-2-4} are novel and have not been reported previously.

\section{Simulation and results}
To verify the correctness of the derived closed-form solution (see Theorem I) and approximations (see Theorem II) numeric simulation was performed. For all the plots (see Fig.1-2), the results obtained with the help of the derived solution  (solid coloured lines) are accompanied with the results derived via numerical integration in \eqref{eq:ABER} (point markers) and simulation (diamond-shaped markers). It is clear that the results accurately match each other. Channel parameters were chosen in such a way to take into consideration: hyper-Rayleigh fading $0.5\leq m<1$ (this is the case of $m=0.5$) and light fading ($m=3.5$ and $m=3$ in Fig. 1 and 2 respectively); and strong/weak dominant component ($K=10$~dB/$-10$~dB). Moreover, the computations were performed both for PSK and lower order QAM modulations, as well as for high-dimensional QAM (actively employed in modern communication standards). The shift $\Delta$ (used to account for integer values of $m$, see Proof of Theorem 1) was set to $10^{-5}$.

\begin{figure}[!t]
\centerline{\includegraphics[width=\columnwidth]{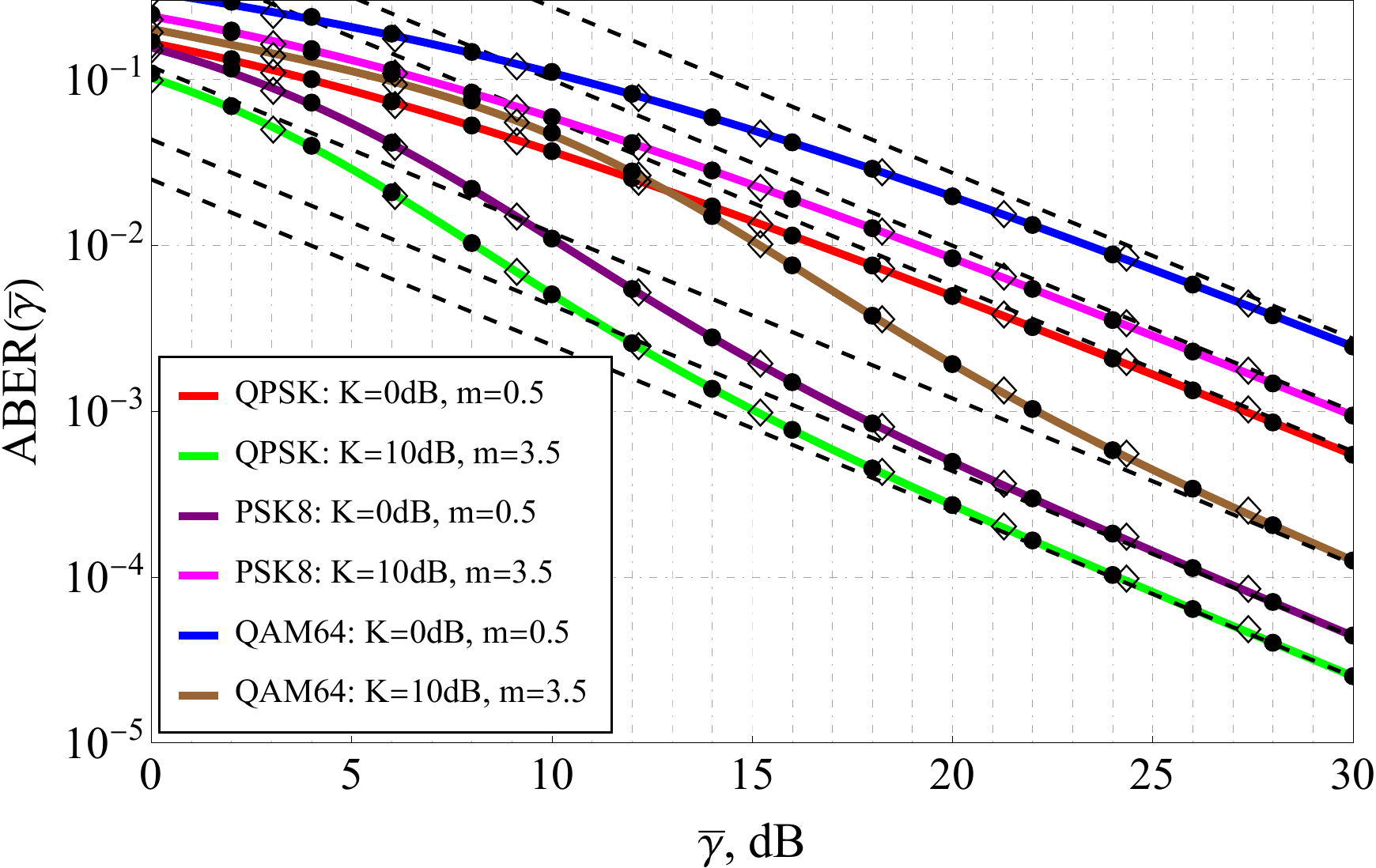}}
\caption{ABER versus $\bar{\gamma }$:solid lines - proposed analytical solution \eqref{thm-1}, point markers - numeric integration in \eqref{eq:ABER}, dashed lines - proposed asymptotic solution \eqref{thm-2-1}, diamond-shaped markers - numeric simulation.}
\label{fig1}
\end{figure}
\begin{figure}[!t]
\centerline{\includegraphics[width=\columnwidth]{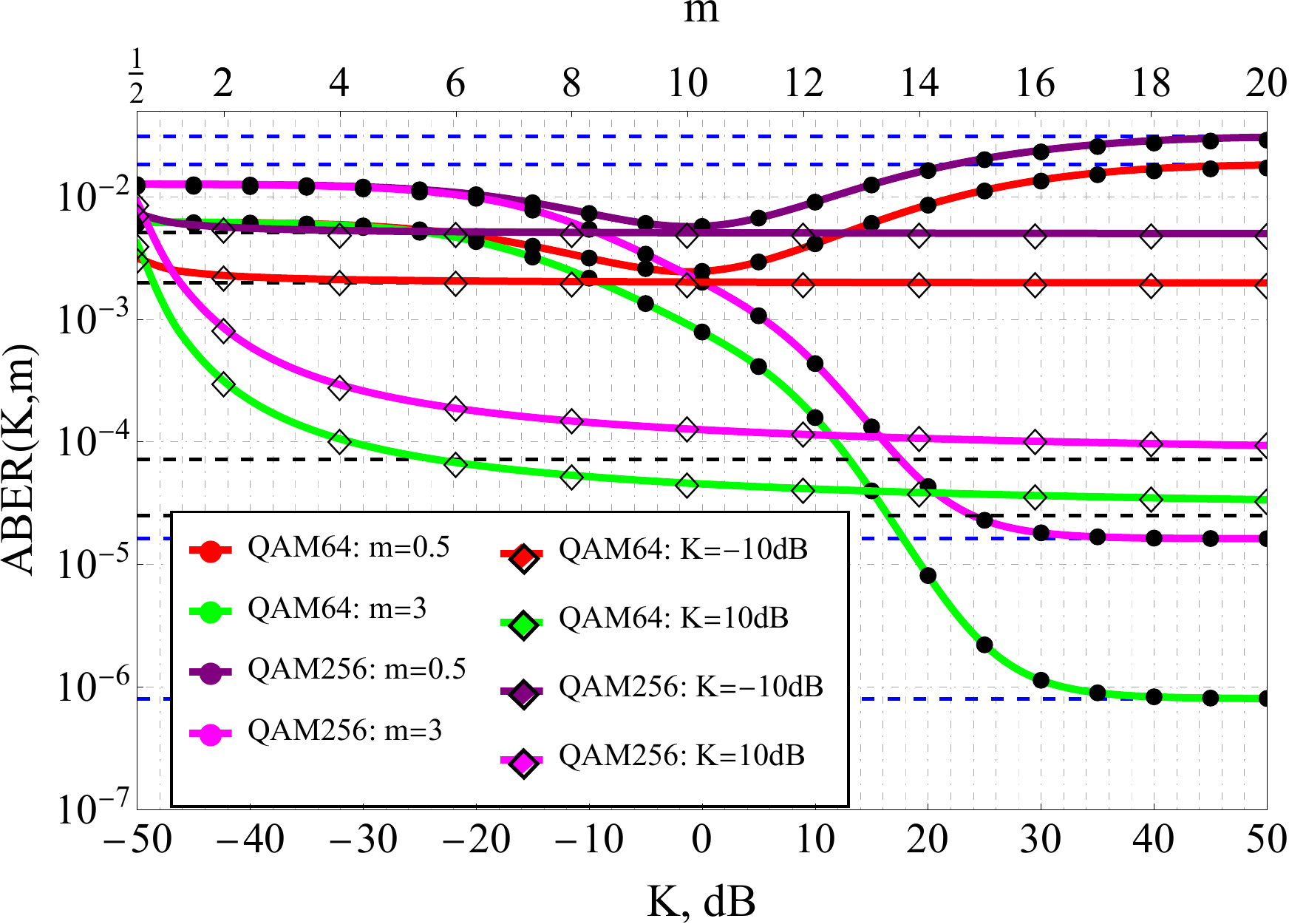}}
\caption{ABER versus $k$ for $\bar{\gamma }=30$~dB: solid lines - proposed analytical solution \eqref{thm-1}, point markers - numeric integration in \eqref{eq:ABER}, dashed blue lines - proposed asymptotic solution for $K\to \infty$ \eqref{thm-2-2}, dashed black lines - proposed asymptotic solution for $m\to \infty$ \eqref{thm-2-3}.}
\label{fig2}
\end{figure}
\begin{table}[!t]
\caption{Relative truncation error for various modulations and fading}
\label{table_example}
\centering
\begin{tabular}{|c|c|c|c|c|}
\addlinespace \hline
$N$ &\makecell{m=0.5\\ QAM-64} &\makecell{ m=2.5\\ QAM-64}&\makecell{m=0.5\\ QAM-1024} &\makecell{m=2.5\\ QAM-1024} \\
\hline
1& $1.64742           \cdot 10^{-2}$& $1.55984           \cdot 10^{-1} $& $2.17328           \cdot 10^{-2}$ & $1.07278           \cdot 10^{-1}$  \\
\hline
2& $5.87588           \cdot 10^{-4}$& $6.005           \cdot 10^{-2}$  &$ 1.01106           \cdot 10^{-3} $& $2.3618           \cdot 10^{-2}$ \\
\hline
3& $3.30798           \cdot 10^{-5}$&$ 1.39152           \cdot 10^{-2}$ &$ 6.71084           \cdot 10^{-5}$ & $2.45898           \cdot 10^{-3}$\\
\hline
4& $2.37544           \cdot 10^{-6}$&$ 4.83533           \cdot 10^{-3}$ &$ 5.80182           \cdot 10^{-6} $& $6.02779           \cdot 10^{-4}$\\
\hline
5& $1.97965           \cdot 10^{-7}$& $2.03268           \cdot 10^{-3}$ & $6.39829           \cdot 10^{-7}$ &$ 2.9732           \cdot 10^{-4}$\\
\hline
\end{tabular}
\end{table}

To study the discrepancy between the closed-form solution and its approximation proposed in Theorem I, an analysis of the relative truncation error $\frac{{\rm ABER_{err}(N,N)}}{\rm ABER}$ was carried out (see Table I), where ${\rm ABER_{err}(N,N)}$ was evaluated with \eqref{cor-1} and ${\rm ABER}$ with numerical integration in \eqref{eq:ABER} assuming $K=5$~dB and $\bar{\gamma }=20$~dB. It is clear that for strong shadowing (i.e., $m=0.5$) only a single term in \eqref{thm-1} is enough (i.e. $L=N=0$) (irrespective of the modulation order), and for moderate $m$ the truncation with $L=N=4$ helps to deliver at least 3-digit accuracy and speed up the calculation (up to 2-4 times, depending on $m$), compared to the numerical integration in~\eqref{eq:ABER}.

For all of the plots, the results obtained with the proposed solution and numeric integration are succeeded by the derived asymptotics: high-SNR asymptotics \eqref{thm-2-1} in Fig. 1 (see dashed black lines), strong/weak dominant component asymptotics \eqref{thm-2-2}/\eqref{thm-2-3} (see dashed blue/black lines in Fig. 2), and light/heavy shadowing asymptotics \eqref{thm-2-4} (see dashed blue/black lines in Fig. 2).

It can be seen (see in Fig. 1) that  for the fdRLoS channel fading deeply impacts the overall system performance: for hyper-Rayleigh case (i.e. $m=0.5, K=0$~dB), even lower-order modulation (QPSK or QAM-4) for high SNR loses about $10$~dB. Moreover, $\bar{\gamma }$ can be easily connected to the relative (to some reference spacing $d_0$) distance between the transmitter and the receiver $d$, accounting for the path loss (via path loss exponent $\alpha$), antenna characteristics and the average channel attenuation ($\chi$), for example, in the following form $\bar{\gamma }=\bar{\gamma }_{\rm tr}+\lg\left(\chi \left(\sfrac{d_0}{d}\right)^{\alpha }\right)$. Here $\bar{\gamma }_{\rm tr}$ (expressed in decibels) is the average SNR at the transmitter output. Thus such a connection with the carried out analysis can help estimate system performance for distance dependant attenuation.

In practice channel parameters are usually unknown and are estimated on-the-go from the measurements, and the inference quality has a crucial effect on the overall system performance. Thus it important to know how great is the impact of the estimated parameters variation on the assumed quality metric. For instance, if the impact is negligible (i.e., asymptotic regime in the corresponding parameter), coarse yet fast inference procedures and algorithms are preferred; otherwise, more complex methods (which are usually slower) are needed.

Fig. 2 demonstrates that scenarios with $K\leq -20$~dB and $K\geq 30$~dB can be assumed as almost asymptotic for any $m$ and constellation size. Moreover, it was found out that in the case of $m<1$ fdRLoS channel exhibits specific performance: $K\to \infty$  actually delivers higher ABER than $K\to 0$; moreover, minimum ABER is observed  when the total power of multipath components is equal to the LoS component (i.e. $K\approx 0$~dB).

It can be observed that the impact of shadowing parameter $m$ in case of weak LoS component is tangible only for $m\leq 2$ irrespective of the modulation order. But for large $K$ asymptotics cannot be reached even with $m=20$ (see Fig. 2). In addition, it is clear that the rate of change of the ABER curve with the increase of $m$ is limited. One can see that the derived asymptotics \eqref{thm-2-1}-\eqref{thm-2-4} excellently describe the ABER floor, which exists due to the shadowed fading nature of the channel.

%

\section{Conclusions}
The letter presents a closed-form and asymptotic analysis for the average bit error rate of a communication system in the presence of the fluctuating double-Rayleigh with the line-of-sight channel. The derived expressions are valid for arbitrary (integer/noninteger) channel parameters and expressed in terms of classical special functions readily accessible in most modern computer algebra systems. The proposed asymptotic bounds cover all possible fading scenarios. All the derived expressions are verified by comparing with the brute-force numerical integration and demonstrated high correspondence with one another.
\appendices{}

\section{Proof of Theorem 1}
\begin{IEEEproof}
To prove Theorem 1, one starts with the fact that the conditional MGF of $\gamma_x$  (i.e. $\mathcal{M}_{\gamma_x}(p|x)=\mathbb{E}\{e^{p\gamma_x}\}$) can be represented as a Laplace transform of the conditional PDF \eqref{eq-pdf-conditional}.
Applying equation (3.35.1.1) (see \cite{Pru92}) and performing simplifications yields:
\begin{IEEEeqnarray}{rCl}\label{eq_thm-1-2}
&&\mathcal{M}_{\gamma_x}(p|x)=-\left(\frac{m}{m+K}\right)^m\frac{(K+1)}{\bar{\gamma }p}\frac{\left(1-\frac{(K+1)}{\bar{\gamma }p}\right)^{m-1}}{\left(1-\frac{m(K+1)}{\bar{\gamma (K+mx)}p}\right)^{m}}
\end{IEEEeqnarray}

It can be noted that the \eqref{eq_thm-1-2} is given in the form of a factorized power-type MGF (see \cite{Gvo21}), thus applying Lemma~2  from \cite{Gvo21} and denoting $\psi_1=\frac{K+1}{\bar{\gamma \delta_{2,j}}}$ and $\psi_2=\psi_1+\frac{K}{m}$ yields:
\begin{IEEEeqnarray}{rCl}\label{eq_thm-1-3}
&&J_Q=\frac{\psi_1}{4}\int_0^{\infty}\frac{F_D^{(2)}\!\left(\frac{1}{2};1\!\!-m,m;2;\frac{\psi_1}{\psi_1+x},\frac{\psi_1}{\psi_2+x}\right)}{(\psi_1+x)^{1-m}(\psi_2+x)^{m}}e^{-x}{\rm d}x,
\end{IEEEeqnarray}
where $F_D^{(2)}(                \cdot )$ is the Lauricella hypergeometric function of two variables \cite{DLMF}. Noticing that $F_D^{(2)}(                \cdot )=F_1(                \cdot )$ (with $F_1(                \cdot )$ being the Appell function) and that its arguments are less that 1 (since $\psi_2>\psi_1$), it can be represented with the convergent series (see equation (16.13.1) in \cite{DLMF}).
Since $F_D^{(2)}$ can be upper-bounded (see (2.15) in \cite{Car66}), yielding a finite majorization, and the summation can be assumed as an integration with the respect to the counting measure, the Fubini's theorem guarantees that the order of the summation and the integration can be interchanged. Thus reorganizing the multipliers, $J_Q$ can be written as:
\begin{IEEEeqnarray}{rCl}\label{eq_thm-1-4}
&&\hspace{-20pt}J_Q\!=\!\sum_{l=0}^{\infty}\sum_{n=0}^{\infty}\!\frac{(\sfrac{1}{2})_{l+n}(1-m)_l(m)_n}{4\psi_1^{-l-n-1}(2)_{l+n}n!l!}     \underbrace{\!\int_0^{\infty}\!\frac{(\psi_1+x)^{m-l-1}}{(\psi_2+x)^{m+n}}e^{-x}{\rm d}x}_{J_1(l,n)}.
\end{IEEEeqnarray}
Applying the relations between the integrands and Meijer G-functions:  $(1+x)^{\alpha }=\frac{1}{\Gamma (\alpha )}G_{1,1}^{\,1,1}\!\left(\left.{\begin{matrix}1+\alpha \\0\end{matrix}}\;\right|\,x \right)$, $e^{-x}=G_{0,1}^{\,1,0}\!\left(\left.{\begin{matrix}\mbox{---}\\0\end{matrix}}\;\right|\,x\right)$ for the case of $\forall m                    \notin                  \mathbb{Z}^{+}_0$ the integral $J_1$ can be evaluated in terms of the extended generalized bivariate Meijer G-function (see equation (13.1) in \cite{Hai92}):
\begin{IEEEeqnarray}{rCl}\label{eq_thm-1-5}
J_1(l,n)&&=\left(\frac{\psi_1}{\psi_2}\right)^m\frac{\psi_1^{-l-1}\psi_2^{-n}}{\Gamma (1-m+l)\Gamma (m+n)}                \times                \IEEEnonumber\\ &&\hspace{-15pt}                \times                 G_{0,1:1,1:1,1}^{1,0:1,1:1,1}\left(
\begin{array}{c}
\mbox{---}\\
0\\
\end{array}\middle\vert
\begin{array}{c}
1-m-n\\
0\\
\end{array}\middle\vert
\begin{array}{c}
m-l \\
0\\
\end{array}\middle\vert
\frac{1}{\psi_2},\frac{1}{\psi_1}
\right).
\end{IEEEeqnarray}

Collecting \eqref{eq_thm-1-3} and \eqref{eq_thm-1-4}, reorganizing the summands and applying the fact that $\frac{(1-m)_l(m)_n}{\Gamma (1-m+l)\Gamma (m+n)}=\frac{\sin{m\pi }}{\pi }$  yields the desired form \eqref{thm-1}.

Finalizing the proof of the first part of the statement, it can be noted that \eqref{eq_thm-1-4} is monotone in $m$ and its rate of change is small enough (see Section~IV). To expand the solution to all possible positive values of $m$ (including integers), one proposes to perform an infinitesimal shift $\Delta$ of the parameter $m$ in case $m                \notin                 \mathbb{Z}^{+}_0 $. Thus the resultant solution is valid for arbitrary values of $m$, as it is demonstrated in Section~IV.

Truncation of the closed-from solution \eqref{thm-1} to $N,L$-terms introduces the error ${\rm err}(L,N)=\!\sum_{\substack{l=L+1\\n=N+1}}^{\infty}\!\!\frac{(\sfrac{1}{2})_{l+n}(1-m)_l(m)_n}{4\psi_1^{-l-n-1}(2)_{l+n}n!l!}J_{1}(l,n)$,
that can be estimated as follows. Integral $J_1(l,n)$ can be upper-bounded by $J_{1}(L,N)$, and since its denominator is increasing in $x$, then $J_{1}(L,N)\leq \int_0^{\infty}\frac{(\psi_1+x)^{m-L-1}}{\psi_2^{m+N}}e^{-x}{\rm d}x=\frac{e^{\psi_1}\Gamma (m-L,\psi_1)}{\psi_2^{m+N}}$,
where $\Gamma (\cdot,              \cdot )$ is the upper incomplete gamma-function. The residual series can be represented as $\sum_{\substack{l=L+1\\n=N+1}}^{\infty}=\sum_{\substack{l=0\\n=0}}^{\infty}-\sum_{l=0}^{L}\sum_{n=0}^{N}$. Assuming that $\psi_1<1$ (needed for convergence), the first series represent Gauss hypergeometric function $\mbox{}_2F_1(\sfrac{1}{2},1;2;\psi_1)$, thus \eqref{cor-1} follows.
\end{IEEEproof}

\section{Proof of Theorem 2}
\begin{IEEEproof}
Assuming that $\bar{\gamma }\to \infty$, integral $\eqref{eq_thm-1-3}$ can be simplified to the following form:
\begin{IEEEeqnarray}{rCl}\label{eq_thm-2-1}
&&J_Q\big|_{\bar{\gamma }\to \infty}                \sim \frac{\psi_1}{4}\int_0^{\infty}\frac{x^{m-1}}{(\psi_1+x)^{1-m}(\frac{K}{m}+x)^{m}}e^{-x}{\rm d}x.
\end{IEEEeqnarray}
It follows from the fact that if $\bar{\gamma }\to \infty$, then $\psi_1\to 0$ and $\psi_2\to \sfrac{k}{m}$, hence the first term of the Taylor series expansion of the $F_D^{(2)}\!\left(\frac{1}{2};1\!\!-m,m;2;\frac{\psi_1}{\psi_1+x},\frac{\psi_1}{\psi_2+x}\right)$ in the vicinity of 0 will be $F_1\!\left(\frac{1}{2};1\!\!-m,m;2;0,0\right)=1$.
Applying the result (13.4.4) from~\cite{DLMF} helps to state that $J_Q\big|_{\bar{\gamma }\to \infty}    \sim \frac{(k+1)}{4\bar{\gamma }\delta_{2,j}}\Gamma (m)U\left(m,1\frac{k}{m}\right),$
where $U(              \cdot )$ is the Tricomi confluent hypergeometric function.


To prove the limiting performance as $K\to \infty$ one can notice that $\sfrac{\psi_1}{K}\to \sfrac{1}{\bar{\gamma }\delta_{2,j}}$ and  $\sfrac{\psi_2}{K}\to \sfrac{1}{\bar{\gamma }\delta_{2,j}}+\sfrac{1}{m}$, thus
\begin{IEEEeqnarray}{rCl}\label{eq_thm-2-3}
&&\hspace{-7pt}J_Q\big|_{K\to \infty}\!              \sim \! \!\int_{0}^{\infty}\frac{\!F_1\!\left(\frac{1}{2};1-m,m;2;1,\frac{m}{\bar{\gamma }\delta_{2,j}+m} \right)}{4\left(\frac{m}{\bar{\gamma }\delta_{2,j}+m}\right)^{-m}}e^{-x}{\rm d}x.
\end{IEEEeqnarray}

Note that in this case Appell function can be simplified, i.e. $F_1\left(\frac{1}{2};1-m,m;2,1,\frac{m}{\bar{\gamma }\delta_{2,j}+m}\right)\to \frac{2}{\sqrt{\pi }}\frac{\Gamma (m+\sfrac{1}{2})}{\Gamma (m+1)}\mbox{}_2F_1\left(\sfrac{1}{2},m;m+1;\frac{m}{m+\bar{\gamma }\delta_{2,j}}\right)$. This yields the desired asymptotics \eqref{thm-2-2}.
It should be specifically pointed out that hereinafter limit and integral operations can be interchanged via dominated convergence theorem since the integrand exhibits a point-wise convergence with the respect to the limiting parameter, and (as it was mentioned in Appendix I) $F_1(   \cdot )$ can be upper-bounded yielding an integrable expression.

For the case of $K\to 0$ one can note that $\psi_2\to \psi_1=\frac{1}{\bar{\gamma }\delta_{2,j}}$. Since the arguments of the Appell function coincide one can make use of the relation (13.4.4)  from \cite{DLMF}, i.e. $F_1(a;b_1,b_2;c;z,z)=\mbox{}_2F_1(a,b_1+b_2;c;z)$. Note that $\mbox{}_2F_1(\sfrac{1}{2},1;2;z)=\frac{2}{z}(1-\sqrt{1-z})$. Then after simplifications
 \begin{IEEEeqnarray}{rCl}\label{eq_thm-2-4}
&&J_Q\big|_{K\to 0}              \sim \frac{1}{2}-\frac{1}{2} \int_{0}^{\infty}\sqrt{\frac{\bar{\gamma }\delta_{2,j} x}{1+\bar{\gamma }\delta_{2,j} x}}e^{-x}{\rm d}x.
\end{IEEEeqnarray}

Evaluating the last integral with the help of equality (13.4.4)~\cite{DLMF} yields \eqref{thm-2-3}. The second form of this asymptotics can be evaluated by relating the Tricomi $U(              \cdot )$ function with the modified Bessel functions (see~\cite{DLMF}).

To find the asymptotic expression for $m\to \infty$, hence $\psi_2\to \psi_1$, one again uses the limiting property of Appell function (see case $K\to 0$) and perform the linear argument transformation of the obtained hypergeometric function $\mbox{}_2F_1(a,b;c;z)=(1-z)^{-b}\mbox{}_2F_1(c-a,b;c;\frac{z}{z-1})$. Noticing that $\mbox{}_2F_1(\frac{3}{2},1;2;-z)=\frac{2}{z}(1-(\sqrt{1+z})^{-1})$ and that $\lim_{m\to\infty}\left(\frac{\psi_1+x}{\psi_1+\frac{K}{m}+x}\right)^m=e^{-\frac{K}{x+\psi_1}}$, the asymptotics can be obtained in the following form
 \begin{IEEEeqnarray}{rCl}\label{eq_thm-2-5}
&&\hspace{-10pt}J_Q\big|_{m\to \infty}      \sim \frac{1}{2} \int_{0}^{\infty}\!\!e^{-\frac{K}{x+\psi_1}}\!\left(1\!-\!\sqrt{\frac{\bar{\gamma }\delta_{2,j} x}{(K+1)+\bar{\gamma }\delta_{2,j} x}}\right)e^{-x}{\rm d}x.
\end{IEEEeqnarray}

Although the solution of the last integral can not be obtained in closed form, it can be easily calculated numerically  via fast and stable procedures.
\end{IEEEproof}

\bibliographystyle{IEEEtran}
\phantomsection\addcontentsline{toc}{section}{\refname}\bibliography{IEEEabrv,BER_fdRLoS}
\vspace{-2cm}

\end{document}